# Is local opposition taking the wind out of the energy transition?*

Federica Daniele**, Guido de Blasio*, and Alessandra Pasquini*
[This draft: 06/02/2024; PRELIMINARY, PLEASE DO NOT CIRCULATE]

## Abstract

Local opposition to the installation of renewable energy sources is a potential threat to the energy transition. Local communities tend to oppose the construction of energy plants due to the associated negative externalities (the so-called "not in my backyard" or NIMBY phenomenon) according to widespread belief, mostly based on anecdotal evidence. Using administrative data on wind turbine installation and electoral outcomes across municipalities located in the South of Italy during 2000-19, we estimate the impact of wind turbines' installation on incumbent regional governments' electoral support during the next elections. Our main findings, derived by a wind-speed based instrumental variable strategy, point in the direction of a mild and not statistically significant electoral backlash for right-wing regional administrations and of a strong and statistically significant positive reinforcement for left-wing regional administrations. Based on our analysis, the hypothesis of an electoral effect of NIMBY type of behavior in connection with the development of wind turbines appears not to be supported by the data.

**JEL Classification**: D72, P18, R12.
**Keywords**: wind turbine installation, electoral outcomes, local opposition.

## Contents

* Bank of Italy. ** OECD and Bank of Italy.

* The views expressed in the paper are those of the authors and do not necessarily reflect those of the institutions they belong to.

## 1. Introduction

In Italy, wind energy total installed capacity stood at 12.1 GW in 2023, corresponding to 27.5% of installed capacity in renewable energy sources (Terna, 2023).[1] However, wind production was less than a third the one produced by the United Kingdom, and less than 20% compared to Germany (OurWorldInData, 2022).

According to the objectives contained in the 2023 proposed revision of *Piano Nazionale per l'Energia e il Clima* (PNIEC), total capacity installed in wind energy plants will need to reach 28.1 GW by 2030 (Ministero dell'Ambiente e della Sicurezza Energetica, 2023), thus requiring an average of newly installed 2 GW p/year between 2023 and 2030, about 5 times as much as the amount of newly installed capacity p/year during the period 2017-22. Importantly, these objectives are likely to be revised upward in light of evolving and increasingly tighter EU targets.

In this work, we ask whether local acceptance (or lack thereof) can be considered a relevant driver of the deployment of wind energy plants. Specifically, we test for the existence of local acceptance as opposed to "Not in My BackYard" (NIMBY) type of behavior triggered by the development of wind turbines. Evidence backing the existence of the latter type of behavior tends to be anecdotal, primarily based on surveys or qualitative data and to overlook a quantification of the political costs directly associated with the development of wind turbines. In this paper, we leverage 20 years of data on wind turbine installations and electoral outcomes referring to regional elections across Italian municipalities to measure the extent of local acceptance with respect to the development of wind turbines. Since 2003, Italian regions ("*Regioni*") are responsible for providing (or denying) the authorization to large energy projects concerning their territories, including wind turbines. The authorization is granted jointly with other authorities, such as the Ministry for the Ecological Transition or the Ministry of Culture. However, as a matter of fact no authorization process can go on without the consent of the regions involved.[2] For this reason, voters are expected to hold regional governments accountable for their stance towards the deployment of renewable – more specifically wind – plants, and the outcome of regional elections to be informative of the potential political costs associated with their construction.

[1] Renewables include solar, wind energy and biofuels, unless otherwise specified.

[2] Since D.lgs. 387/2003 the regions have the power to implement a legislation favoring, in their territories, a faster deployment of renewable plants (Daniele et al., 2022) as a further proof of their central role in shaping the development of renewable plants.

Following Bayulgen et al. (2021) and Urpelainen and Tianbo Zhang (2022), we acknowledge that the political consequences of favoring the development of wind turbines might differ based on party identity. This exercise is made easier by the two-party (or two-coalition) nature of the Italian political system (both at the national and at the regional level) characterized, until 2018 national elections, by a center-left and a center-right coalition, with traditionally opposing views concerning the urgency to adopt policies that reduce carbon emissions.[3] In a similar vein to Bayulgen et al. (2021), in our baseline specification, we regress the share of votes obtained in a given municipality by the incumbent coalition in regional elections on a dummy variable equal to one if in that same municipality a wind turbine was developed during the course of the previous mandate. This dummy variable is further interacted with another dummy capturing whether the incumbent coalition was left-wing. The endogeneity of wind turbine development is tackled through the use of wind speed as an instrument. A set of municipality and time-varying controls are further included to address remaining concerns of omitted variable bias. The estimation is carried on Southern regions exclusively. This allows minimizing sample heterogeneity given the wide North-South divide in Italy, without sacrificing the representativeness of the analysis given that Southern regions accounted for 96% of total installed capacity in wind plants as of 2020.

Our baseline results indicate that the degree of local acceptance of renewable projects heavily depends on the political orientation of the regional government at the time of projects' development. More specifically, we find robust evidence for an electoral premium associated with wind turbine development in municipalities located in regions with a left-wing incumbent coalition. In Italy, left-wing regional governments are empirically associated with more widespread environmental concerns among the local population. Hence, our finding is in line with theories of positive reinforcement, according to which the development of a wind turbine likely induces a positive reaction among voters who backed the incumbent administration also owing to their pro-environment/renewables stance. Conversely, we do not find convincing evidence of the existence of an electoral cost associated with wind turbines development when the incumbent regional government is right-wing. The associated estimated impact is always negative, and larger in absolute value when the development concerns large plants, but never statistically significant. These results are robust to the addition of a large set of covariates and do not seem to be sensitive to the presence of spatial spillovers.

[3] Since 2013, the Five Star Movement has risen as third political party and coalition in Italian political landscape. While the party has come to power at the national level in 2018, they have been so far struggling at conquering seats and a clear majority in regional elections, which continue up to now being characterized essentially by a two-coalition system.

The impact of renewable plants development on electoral outcomes is heterogeneous. We test whether local characteristics that might cause higher social costs from renewable plants development (e.g., in terms of decline in land values or tourism flows) are associated with stronger electoral backlash. Electoral backlash for right-leaning incumbent regional governments is stronger (albeit never statistically significant) in municipalities with high house prices/tourist penetration compared to municipalities where these are low. Furthermore, we find electoral backlash/positive reinforcement to be stronger in the second half of the sample (after 2013). This last heterogeneity result may be due to the overall increase in sensitivity towards climate change over time. Finally, both electoral backlash and positive reinforcement appear to be stronger in magnitude if wind turbines were developed during the second half of the previous mandate, in line with the fact that in this case they might be more salient in voters' memory.

Finally, we discuss the external validity of our findings. Our main concern is that the estimated local average treatment effect (LATE) cannot capture the impact of wind turbine development on electoral outcomes in municipalities where development/no development occurs regardless of wind conditions and the treatment effect in these areas is different. We acknowledge the external validity limitations from which the chosen identification strategy potentially suffers and lay out a strategy to test for whether these concerns are justified. Based on the updated strategy, we are able to confirm the external validity of the absence of electoral backlash, but not the positive reinforcement for left-wing incumbent regional administrations.

The remainder of the paper is articulated as follows: Section 2 provides a literature review, Section 3 describes the data, Section 4 goes into the details of the empirical analysis, Section 5 concludes.

## 2. Literature review

There is an emerging literature on local opposition to the development of wind projects and on its determinants,[4] mostly based on descriptive analysis of survey data. The bulk of this literature agrees that local opposition exists (Bigerna and Polinori (2015), Jones and Eiser (2009), Jones and Eiser (2010), Kontogianni et al. (2014), Maggi et al. (2015)), with the visual impact and noise generated by wind turbines being important drivers of local opposition (Wolsink (2000), Kontogianni et al. (2014), Hoen et al. (2019)). Few papers, however, suggest that it is marginal or does not exist (Swofford and Slattery (2010), Theron et al. (2011)).

Effective communication and the involvement of local communities can effectively soften local opposition and drive social acceptance (Fergen and Jacquet (2016), Hoen et al. (2019), Kontogianni et al. (2014)). In Italy, based on a series of interviews held with local stakeholders and authorities in three areas of the region of Puglia where wind turbines were developed, Maggi et al. (2015) detect a sense of discontent of the local community towards local authorities and developers due to their little involvement and the perceived lack of transparency over the authorization and siting procedures. Devine-Wright (2005), Petrova (2013), Ellis and Ferraro (2016) all suggest that political and/or economic participation to the development and management of wind plants should be used to overcome local opposition.[5]

The conclusions based on survey data should be taken with caution (Stratton, 2015). Besides the limitations normally characterizing survey data[6], several of the reviewed studies have a low response rate. Additionally, it is reasonable to think that the individuals most unsatisfied and angry about the development of the projects are those more likely to answer to the questionnaire. Therefore, the data may overestimate the presence of local opposition.

Another strand of the literature – primarily North American-based – tries to detect empirically local opposition by using quasi-experimental methods. The common approach is to exploit electoral outcomes as a proxy for local opposition. The underlying idea is that, in presence of local opposition, local communities will penalize those that are (or that are believed to be) responsible for the development of the wind turbines. Bayulgen et al. (2021) estimate the impact, at the precinct level, of the construction of one additional wind turbine on the incumbent share at the Minnesota House of Representatives' next term elections during the time span of 2006-2018. In their work, the potential endogeneity in the siting of wind turbines is instrumented for with wind speed and the log of precincts' surface. One additional turbine is found to increase the incumbent share by 2.5 percentage points, suggesting that the development of wind turbines is locally accepted (and rewarded). Although similar in magnitude, the effect is more precisely estimated for the Republican Party, a result that they attribute to the potentially improved economic outcomes (e.g., new employment) triggered by wind energy investments. Urpelainen and Tianbo Zhang (2022) estimate the impact of the construction of wind turbines, at electoral district level, on the Democratic Party share at the United States House of Representatives elections between 2003 and 2012. Using wind speed as an instrument for wind turbine construction, wind turbine development is shown to have a positive effect on the Democratic Party share. In further analysis, they employ as an outcome the share of the incumbent candidate checking the heterogeneity of the result with respect to candidate's affiliation party. The effect is shown to be negative when the incumbent party is Republican while positive when it is Democratic (although it is never significant). They refer to these results as evidence in favor of the "positive reinforcement" hypothesis (opposed to the "local opposition" one). Starting from the premise that the Democratic Party is expected by its constituency to favor renewable energy sources deployment, according to this hypothesis its voters are pleased when wind turbine deployment took place when the Party was in power and they reward them during the following election. A shortcoming of both these studies is that the average treatment effect is estimated based exclusively on areas where wind turbines development took place. Hence, neither of them allows to detect the potential electoral backlash characterizing areas where no development occurred because of local opposition. Finally, Stokes (2016) analyzes the impact of a national policy introduced by the Liberal Party in Canada that favored the development of wind turbines and detects a reduction of support to the Liberal Party between 7 and 10 percentage points in areas where wind turbines were developed.

[4] See Devine-Wright (2005), Petrova (2013), Ellis and Ferraro (2016) for some literature reviews.

[5] Examples of economic participation include providing compensations to local communities as well as the participation to renewable energy generation by means of the establishment of energy communities.

[6] If the questions are not properly phrased the respondents may misunderstand them and their responses may reflect excessively their perceptions. The respondents may answer what is perceived as the most beneficial or the most socially accepted opinion rather than being honest.

Based on the literature, the development of wind turbines has significant electoral consequences. Moreover, it has the potential to trigger two-way reactions. On the one hand, it can cause electoral backlash if voters' preferences put greater weight on the associated costs vs. benefits. On the other, it can induce positive reactions among voters if their preferences are such that the benefits outweigh the costs. In a two-party political system equilibrium, the two opposing parties will develop political platforms serving either the green or less green constituency. Then, if the development of wind turbines is favored by the party (or coalition) serving the less green constituency, we would expect the electoral backlash mechanism to dominate. If, vice-versa, the development of wind turbines is favored by the party (or coalition) serving the green constituency, a sentiment of approval or "positive reinforcement" among voters should prevail (Urpelainen and Tianbo Zhang, 2022).

## 3. Data

To conduct our analysis we build a panel dataset covering most southern regions and spanning the period 2000-2020.[7] We focus on southern regions as these are the areas mostly interested by the installation of wind turbines.[8] The geographical distribution of wind turbines reflects indeed that of wind speed, with 96% of total capacity in 2020 installed in southern regions. Furthermore, the choice of focusing on southern regions reduces substantially the heterogeneity in our sample, thus simplifying the task of deriving causal inference.

We build the dataset using different sources, which we describe below in greater detail.

### *3.1 Electoral outcomes*

Data on electoral outcomes come from two distinct datasets. First, the *Anagrafe degli Amministratori Locali e Regionali*, managed by the Italian Ministry of Internal Affairs, containing information on elected candidates. Second, *Eligendo*, also managed by the Ministry of Internal Affairs, containing information on electoral outcomes at the municipality level for all types of Italian elections (i.e., the number of eligible and actual voters, the number of votes each candidate received, the number of invalid votes, etc.).

We focus on regional elections since the *Regioni* are the administrative level determining the policy orientation on the subject of renewable plants development. There are 41 regional elections occurring during the period of interest in continental Southern regions.[9] Up to the regional administrative level, the Italian political system is essentially a two-party/coalition one. We proceed by manually searching for the center-right and center-left regional presidential candidate during each electoral run. Further, we define the incumbent coalition as the coalition assigned to the candidate that has been in power during the previous mandate. Finally, we calculate for each election and municipality the share of votes obtained by the incumbent coalition in that municipality.

[7] Due to data availability, when specified, we focus over the period 2006-2020 or 2005-2020.

[8] The share of municipalities/year featuring the development of at least a wind turbine is less than 1% for municipalities located in the Centre-North, while it is 6% for municipalities located in the South.

[9] Electoral outcomes on Sicily and Sardinia are missing due to the partial fiscal autonomy of these regions, who are not obliged to communicate data on electoral outcomes to the Ministry of Internal Affairs.

The supporters of wind turbine development are more likely to be left-oriented (Bigerna and Polinori, 2015). To provide additional evidence on the relation between the color of political coalitions and citizens' attitude towards the environment we follow Hoffmann et al. (2022) in using Eurobarometer data to construct a region/year varying measure of environmental concerns among the population. Next, we test whether in the region/years of electoral runs with a left-wing elected regional president environmental concerns are more widespread compared to those with a right-wing elected politician. Eurobarometer data suggest that in region/years of electoral runs with a left-wing elected politician the share of voters ranking environmental issues among the most relevant at a national level is 4.4%, against 3.5% when the elected regional government was right wing. This evidence suggests that Italian regional elections are a suitable ground to test the theories of electoral backlash vs. positive reinforcement.

### *3.2 Wind turbine*

We combine data on electoral outcomes with data on the development of wind turbines provided by the *Gestore Servizi Energetici S.p.A.* (or GSE S.p.A.), a publicly owned company in charge of promoting and supporting the deployment of renewable energy sources in Italy. GSE S.p.A. data cover all renewable energy plants conditional on having received a financial subsidy (see Fig.1 for the evolution of wind installed capacity during 2007-2019). For wind power, that meant approximately 91% of total capacity installed in Italy at the end of 2020.[10] The GSE S.p.A. has good coverage of renewable plants in Italy also from a size distribution point of view. The share of energy installed in plants above 10 MW was in 2020 89% of the total against 90% according to Terna, while the share installed in plants between 200 kW and 10 MW was 8.4% against 8% according to Terna. The observations are at wind energy plant level and the database include information on the location, the power, the year and the month the plant was built. See also Daniele et al. (2022) for further details on the GSE S.p.A. database.

[10] The total capacity installed amounted to 10 GW according to GSE S.p.A. database against 10.9 GW according to Terna, the Italian energy transmission operator.

**Figure 1: Diffusion of wind energy across Italian municipalities in 2007, 2013 and 2019**

(a) 2007

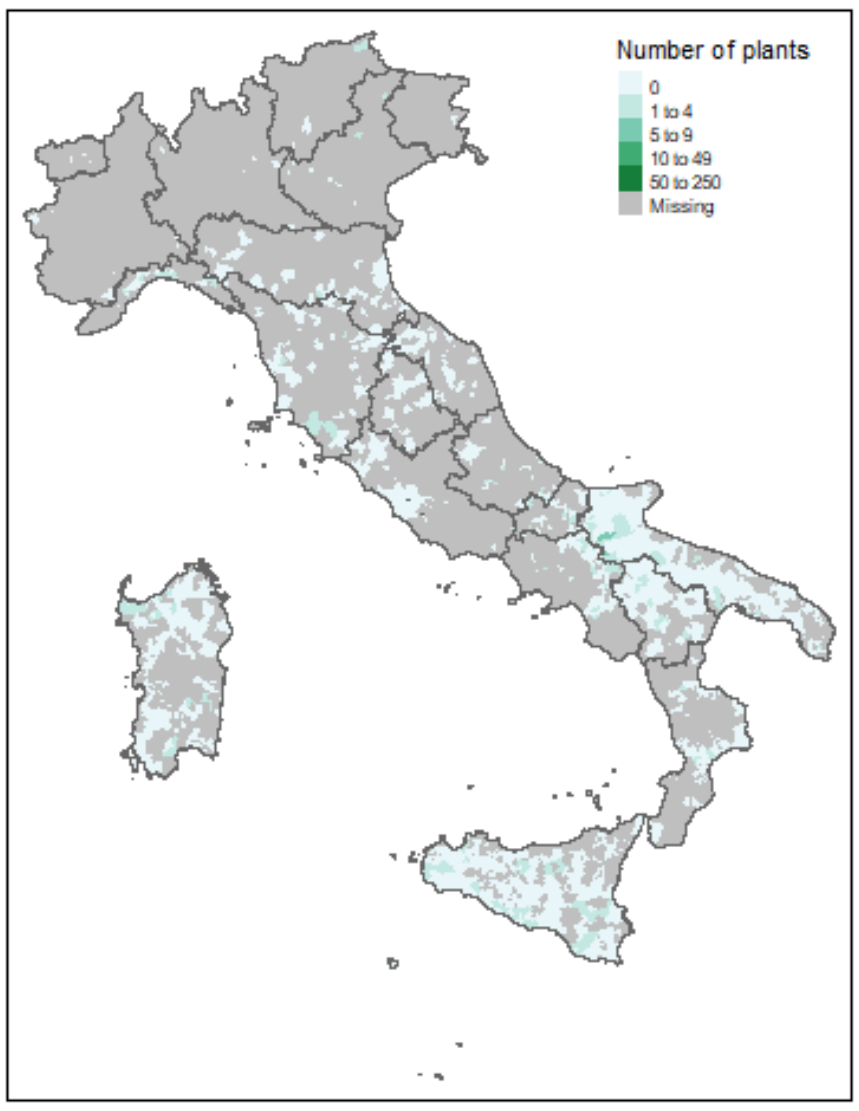

(b) 2013

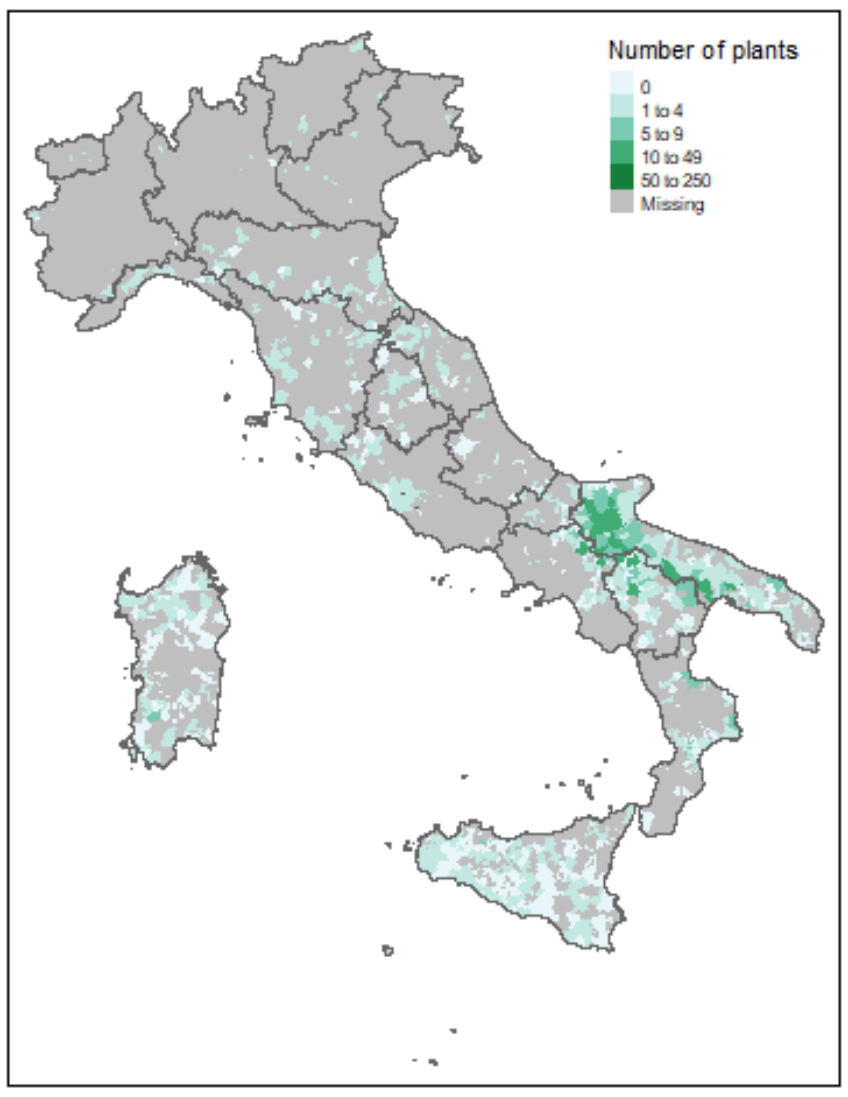

(c) 2019

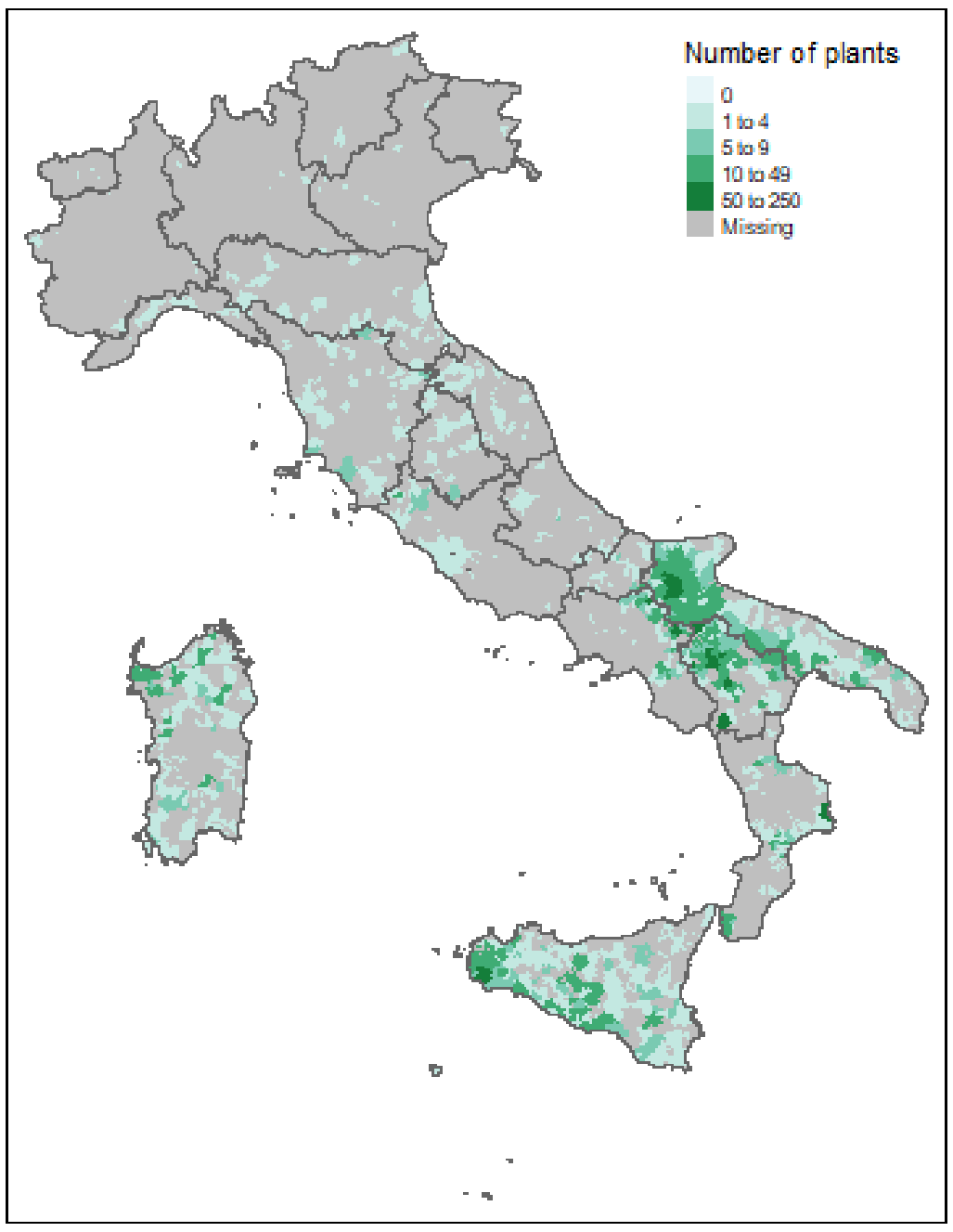

Source: Gestore Servizi Energetici S.p.A.

*3.3 Other data sources*

In the analysis, we further employ data coming from other sources as listed in Table 1 (see Table 2 for descriptive statistics). Wind speed in the municipality at 50 meters of height was used as an instrument in the empirical strategy. Data on population, income, municipality's surface and elevation where used as control variables. We use in the heterogeneity analysis data on real estate prices, tourism intensity, population density and waste sorting.

**Table 1: Data sources employed in the analysis as controls**

| Variable | Time varying | Year(s) | Source |
|---|---|---|---|
| Population | Yes | 2000-2020 | Istat |
| Per capita declared income | Yes | 2000-2020 | Istat |
| Surface of municipality | No | - | Istat |
| Altitude of municipality | No | - | Istat |
| Wind speed | No | - | *GSE* |
| Average real estate residential price | No | 2005 | OMI, Ministry of Economics and Finance |
| Number of accommodation facilities | No | 2004 | Istat |
| Waste sorting | No | 2003 | Ispra |

Notes: wind speed is measured as the annual average wind speed 50 meters above the level of the sea. The number of beds reports the number of beds in accommodation facilities. All variables are defined at municipality level with the exception of waste sorting, which is defined at province level.

**Table 2: Descriptive statistics**

| Variable | N | Mean | Std. dev. | Min | Max |
|---|---|---|---|---|---|
| Share incumbent (%) | 6,504 | 44 | 18 | 0 | 93 |
| At least one wind turbine (dummy) | 6,504 | 0.07 | 0.26 | 0 | 1 |
| Left-wing incumbent (dummy) | 6,504 | 0.62 | 0.49 | 0 | 1 |
| Population | 6,504 | 8314 | 31465 | 85 | 992779 |
| Average yearly income (EUR) | 6,504 | 11348 | 2383 | 5744 | 22702 |
| Surface (km2) | 6,504 | 39.73 | 47.87 | 0.12 | 593.93 |
| Median altitude (m) | 6,504 | 428 | 343 | 2 | 1741 |
| Wind speed | 6,504 | 5.26 | 0.77 | 3.24 | 7.77 |
| House price (EUR p/m2) | 6,406 | 666 | 409 | 245 | 7349 |
| Number of accommodation facilities | 6,504 | 5 | 14 | 0 | 271 |
| Waste sorting (%) | 6,474 | 8.38 | 3.32 | 2.62 | 19.19 |

Notes: averages during 2000-2020. Only municipalities located in the South of Italy.

## 4. Empirical analysis

Our goal is to test whether the development of wind turbines triggers a systematic shift in voter preferences depending on the affiliation of incumbent regional governments. We use municipalities as the unit of our analysis and estimate the following equation:

$$y_{it} = \alpha + \gamma_t + \delta \boldsymbol{X}_{it} + \beta_1 D_{it} + \beta_2 D_{it} * L_t + \varepsilon_{it} \quad \text{(I)}$$

where $y_{it}$ measures the share obtained by the incumbent coalition in municipality $i$ during election $t$ occurring in $i$'s region, $D_{it}$ is a dummy variable taking value 1 if at least one wind turbine was installed in the municipality during the mandate preceding election $t$. $L_t$ is a dummy variable taking value 1 if the incumbent coalition in election $t$ had a center-left affiliation. Finally, $\gamma_t$ are election fixed effects and $\boldsymbol{X}_{it}$ is a set of control variables referring to the municipality in the election year.

In this specification, the coefficients $\beta_1$ and $\beta_2$ measure the correlation between wind turbine installation and the share obtained by the incumbent coalition in each municipality. If the timing and location of wind turbine were exogenous (i.e., randomly distributed with respect to the share of the incumbent) the two coefficients would measure the impact of a wind turbines installation on the share of the incumbent coalition differentiating by political coalition. Nonetheless, the timing and location of wind turbines development is unlikely exogenous. For example, consider the case of the incumbent coalition experiencing a fall in local popularity during the mandate. The candidate may decide to support/oppose the installation of wind turbines in order to gain the votes of supporters/opponents. Therefore, electoral expectations (likely correlated with electoral outcomes) influence the probability of wind turbine installation, introducing an endogeneity issue.

### *4.1 Identification strategy*

In order to address the endogeneity issue in our setting, we use an instrumental variables approach. Following Bayulgen et al. (2021) and Urpelainen and Tianbo Zhang (2022), we use average wind speed measured at 50 meters over the sea as an instrumental variable. Thus, we estimate the impact of a wind turbine being activated when its installation is completely determined by the presence of wind in the area. A valid instrument has to satisfy two requirements: 1. Being highly correlated with the treatment (i.e., being a strong instrument).

Wind speed satisfies this condition, as we document below, being arguably an essential determinant of wind turbine installation. 2. Affecting the outcome exclusively through the treatment (i.e., satisfying the exclusion restriction) and not being spuriously correlated with the outcome. While the exclusion restriction is satisfied under the plausible hypothesis that voters' decisions are not influenced by how windy their municipality is, we provide below some tests to support the credibility of the second part of this requirement.

Given that our endogenous treatment is a binary variable we follow a three step approach (for a discussion of its advantages see Wooldridge (2002) and Adams et al. (2009)). First, we use a probit model to calculate the predicted probability of treatment. Second, we run an OLS estimation, as in the standard first stage of a 2 Stages Least Squares regression (2SLS), but using the predicted probability in lieu of the instrument. Third, we run the second stage of the 2SLS as usual (see the description of probit-2SLS in Cerulli, 2014).

### *4.2 Results*

Table 3 contains the baseline estimation results. In columns (1) - (3) we report the results of an OLS estimation of equation (I) without control variables and respectively with, and without, the interaction between the treatment and the left incumbent dummy. The installation of at least a wind turbine is significantly correlated with a higher incumbent share if the latter belongs to the center-left coalition, while it is negatively and not significantly correlated with the incumbent share if the incumbent belongs to the center-right coalition. Columns (4) - (5) report the results of the probit-2SLS estimation. In column (5) the effect is split between center-left and center-right coalitions.

**Table 3: Incumbent coalition share on wind turbine installation: baseline results**

| | Incumbent coalition share | | | | |
|---|---|---|---|---|---|
| | (1) | (2) | (3) | (4) | (5) |
| | OLS | OLS | OLS | Probit-IV | Probit-IV |
| Treat | 0.53 | -3.31** | -1.83 | 6.78** | |
| | (0.56) | (1.45) | (1.13) | (2.70) | |
| Treat*Left | | 5.74*** | 3.31** | | |
| | | (1.59) | (1.30) | | |
| LATE for right incumbent | | | | | -0.66 |
| | | | | | (4.15) |
| LATE for left incumbent | | | | | 8.62*** |
| | | | | | (2.45) |
| Observations | 8,152 | 8,152 | 8,152 | 6,505 | 6,504 |
| Election FE | YES | YES | YES | YES | YES |
| Controls | NO | NO | YES | YES | YES |
| 1st Stage z-stat[(a)] | - | | - | 15.20 | 15.20 |
| | | | | (0) | (0) |
| Std errors | Robust | | Robust | Robust | Bootstrapped |

Notes: standard errors in parentheses. Significance level: *** p<0.01, ** p<0.05, * p<0.1. "Treat" is a dummy taking value 1 if at least a wind turbine was installed in the municipality in the previous mandate. "Left" is a dummy taking value 1 if the incumbent belong to the center-left coalition. The Probit-IV uses as instrument the annual average wind speed 50 meters above the level of the sea. Control variables include log of average income, log of population (municipality and time-varying), municipality surface and median altitude (time-invariant). The lower number of observations in the probit-IV analysis is due to the software exclusion of some observations due to failing in prediction during the first step. Most of them refers to elections of year 2000 and the OLS estimation is fairly robust to the exclusion of the same observations (see Appendix A). This suggests this is not a major concern. [(a)] This is the 1st stage test statistics z corresponding to the instrument's coefficient with the corresponding p-value in parentheses.

When moving from the OLS to the probit-2SLS model the estimated coefficient strongly increases in magnitude. This suggests the presence of downward bias in the OLS estimation, such as the one triggered by a situation where incumbent coalitions favor the development of wind turbines in municipalities/years where/when they are less likely to achieve a good result during next term elections. Overall, our preferred estimation (column (5)) suggests that the sentiment of "positive reinforcement" among voters dominates the one of "electoral backlash" when the development of wind turbines occurs in municipality/years subject to the government of a regional center-left coalition - typically more likely to serve green constituencies. The sentiment of electoral backlash appears to be stronger in municipality/years subject to the government of a regional center-right coalition, but not enough to dominate over the one of positive reinforcement and impose itself as statistically significant determinant of electoral outcomes of center-right regional governments.

These results are in line with the literature. They are qualitatively in line with the results by Urpelainen and Tianbo Zhang (2022), who find evidence in favor of a positive reinforcement mechanism for the Democratic Party (US equivalent of Italian's center-left coalition) and an un-precisely estimated negative effect on the share of the Republican Party (US equivalent of Italian's center-right coalition). The coefficients reported in Table 3 appear larger than in Urpelainen and Tianbo Zhang (2022). However, this may be due to the different definition of treatment in their study (i.e., cumulative capacity). Our results echo also the finding in Bigerna and Polinori (2015), who find – based on a survey conducted in the region of Umbria - that survey respondents supporting the development of wind turbines are more likely to be left-oriented.

### *4.3 Robustness*

The first assumption we make when using an instrumental variable approach is the inclusion restriction. The instrument has to be a strong predictor of the treatment. This is likely to be the case in our framework. Wind speed is arguably a relevant and essential determinant in wind turbine installation. Moreover, the first stage statistics displayed in the previous section suggest that the instrument is strong enough. The second assumption we make is the absence of spurious correlation and the exclusion restriction, namely we assume that the incumbent coalition share is affected by wind speed exclusively through an increase in the probability of wind turbine installation during the previous term. This assumption would be violated if voters' choice is affected by how windy their municipality is. This is hardly believable. However, this assumption may be violated if, e.g., municipalities with higher wind speed are mountainous and/or peripheral and voters' dissatisfaction with politicians is higher in these areas, perhaps because of longstanding economic decline. To overcome this possible violation, the addition of municipality fixed effects would be desirable. However, this cannot be done when employing a time-invariant instrument at the municipality-level. To compensate for this, we added to the estimation a large battery of controls. We controlled for a set of variables that could potentially be influenced by wind speed and influence voters' electoral decisions, such as average income and population. It is impossible to test whether controlling for this set of variables is enough to remove all spurious correlation. Nonetheless, we can check for whether the absence of further omitted variables is plausible in our context. In order to do so, we can add other covariates to the regression model and check if the estimated treatment effect changes significantly. If any of those variable is the source of spurious correlation between the instrument and the outcome we expect the estimated effect to change significantly. If, on the contrary, it does not change, our concern will be reduced. In Table 4 we report the estimated treatment effect according to our standard model (column 1) and when the additional covariates are included (column 2). The estimated effects are not significantly different, thus re-assuring us on the plausibility of the identifying assumptions.

**Table 4: Incumbent coalition share on wind turbine installation: testing identifying assumptions**

| | Incumbent coalition share | | |
|---|---|---|---|
| | (1)<br>Standard estimation | (2)<br>Additional covariates added | (3)<br>Municipalities surface <50 Km$^2$ |
| LATE for right incumbent | -1.50 | -.88 | 1.48 |
| | (4.58) | (4.44) | (7.00) |
| LATE for left incumbent | 8.21*** | 7.58*** | 20.93*** |
| | (2.99) | (2.75) | (6.18) |
| Observations | 4,720 | 4,720 | 4,859 |
| Years | 2006-2020 | 2006-2020 | 2000-2020 |
| Election FE | YES | YES | YES |
| Controls | YES | YES | YES |
| Additional controls | NO | YES | NO |
| 1st Stage z-stat[(a)] | 9.98 | 9.70 | 7.48 |
| | (0) | (0) | (0) |

Notes: Bootstrapped standard errors in parentheses. Significance level: *** p<0.01, ** p<0.05, * p<0.1. The Probit-IV uses as instrument the annual average wind speed 50 meters above the level of the sea. Standard control variables include log of average income, log of population (municipality and time-varying), municipality surface and median altitude (time-invariant). Additional control variables include a dummy identifying municipalities whose province recycled more than 10% of their wastes, a dummy identifying rural municipality, a dummy identifying municipalities whose average residential real estate price in 2005 was higher than the South-Italy median, a dummy identifying municipalities with at least one accommodation facility in 2004 and a dummy identifying municipalities with a number of beds in accommodation facilities in 2004 higher than the South-Italy median. The analysis in (1) and (2) focus on the period 2006-2020 as some of the covariates, potentially affected by the treatment, are not observed earlier than 2005. [(a)] This is the 1st stage test statistics z corresponding to the instrument's coefficient with the corresponding p-value in parentheses.

Do the data reveal the existence of spillover effects? If positive reinforcement is the result of a climate-friendly attitude by voters, the installation of a wind turbine in a given municipality may increase the incumbent coalition share in other municipalities located in the same region, thus inducing positive spillovers on control units and reducing the estimated positive impact (for center-left incumbent coalitions). We would expect this phenomenon to be stronger in those areas where voters are more climate friendly. However, in Table 5a below, we do not find evidence of stronger positive reinforcement in areas where there is higher ecological sensitivity, thus suggesting that this kind of spillovers do not seem to be a source of bias in our results. Alternatively, positive reinforcement may be due to the fact that the development of a wind turbine benefits economic activity, e.g., in terms of employment, infrastructure, etc. These benefits are likely to diffuse spatially around the location where the wind turbine is located, thereby impacting positively incumbent coalitions' electoral outcomes also beyond treated municipalities and reducing the overall estimated positive impact for center-left incumbent coalitions. This type of spillovers is more likely to materialize in small municipalities, where the economic impact of wind turbines is more likely to spread beyond the treated municipality borders. If spillovers are a valid reason of concern, the impact estimated on municipalities with a small surface should be lower than the baseline estimated impact. We test whether this is the case by repeating our estimation on municipalities whose surface is equal or less than 50 km$^2$ (column 3 of Table 4). The impact estimated on small municipalities is higher, thereby ruling out spillovers as a reason of concern in our setting.

### *4.4 Heterogeneity*

Most of the literature on local opposition to renewable energy plants has highlighted the existence of heterogeneity with respect to local and/or individual characteristics (Devine-Wright (2005), Jones and Eiser (2009), Jones and Eiser (2010), Kontogianni et al. (2014), Petrova (2013), Swofford and Slattery (2010), Theron et al. (2011), Hoen et al. (2019), Ellis and Ferraro (2016)). We contribute to this literature by investigating whether the sentiments of electoral backlash/positive reinforcement change with respect to some local characteristics that we deem of interest. Hence, we augment equation (I) in the following manner:

$$y_{it} = \alpha + \gamma_t + \delta \boldsymbol{X}_{it} + \theta_1 H_{it} + \theta_2 D_{it} + \theta_3 D_{it} * L_t + \theta_4 H_{it} * D_{it} + \theta_5 H_{it} * L_t + \theta_6 H_{it} * D_{it} * L_t + \varepsilon_{it} \quad \text{(II)}$$

where $H_{it}$ is a dummy variable constructed based on a given local characteristic of interest. The probit-2SLS approach is used to calculate the average treatment effect for different observations corresponding to different combinations of $H_{it}$ and $L_t$ values. Table 5a – 5b show the estimated LATE.

**Table 5a: Incumbent coalition share on wind turbine installation: heterogeneity results**

| | (1) Rural | | (2) High ecological sensitivity | | (3) High real estate cost[(a)] | | (4) Touristic | |
|---|---|---|---|---|---|---|---|---|
| | 0 | 1 | 0 | 1 | 0 | 1 | 0 | 1 |
| C-Right | 0.19 | 1.58 | -3.28 | -2.33 | 2.36 | -0.52 | 2.21 | -3.23 |
| | (5.08) | (4.55) | (4.67) | (5.43) | (4.95) | (4.56) | (4.92) | (5.08) |
| C-Left | 4.98* | 6.37*** | 7.26*** | 8.21*** | 10.20*** | 7.32*** | 11.27*** | 5.83** |
| | (3.46) | (2.72) | (2.97) | (3.25) | (3.51) | (3.21) | (2.94) | (2.78) |
| Observations | 6,504 | | 6,504 | | 4,720 | | 5,895 | |
| Years | 2000-2020 | | 2000-2020 | | 2006-2020 | | 2005-2020 | |
| Election FE | YES | | YES | | YES | | YES | |
| Controls | YES | | YES | | YES | | YES | |
| 1st stage z-stat(b) | 14.67 | | 15.18 | | 14.22 | | 15.25 | |
| | (0) | | (0) | | (0) | | (0) | |

Notes: The values reported are the average treatment effects. Bootstrapped standard errors in parentheses. Significance level: *** p<0.01, ** p<0.05, * p<0.1. The estimating equation is (II) where the treatment is instrumented with the annual average wind speed 50 meters above the level of the sea in a probit-2SLS approach. Control variables include log of average income, log of population (municipality and time-varying), municipality surface and median altitude (time-invariant). Rural municipality is a dummy taking value 1 if the municipality average population density is below 150 inhabitants p/squared kilometer, and 0 otherwise. "High ecological sensitivity" takes value 1 if in the corresponding province recycling share is above 10%, and 0 otherwise. (a) 9 observations are excluded because of missing values for the real-estate cost. (a) This is the 1st stage test statistics z corresponding to the instrument's coefficient with the corresponding p-value in parentheses.

In column (1) we check the heterogeneity with respect to the municipality being rural. Following the guidelines of the European Commission, a territory is defined as rural if population density is below the threshold of 150 inhabitants per squared kilometer. According to this definition two thirds of municipalities in the sample qualify as rural. The results highlight a stronger positive reinforcement in rural municipalities. This may be due to the fact that these municipalities are less densely populated, henceforth the local communities will perceive less the costs of wind turbines.

In column (2) we analyze whether voters' response to wind turbine development varies depending on the local recycling share, as a proxy for ecological sensitivity. We classify municipalities as having "high ecological sensitivity" if the province recycling share is above 10%, halfway between 0% and 20%, the highest value. Our findings suggest that in municipalities with high ecological sensitivity the positive reinforcement is slightly strengthened while the (small and non-significant) negative backlash is slightly reduced. This is in line with the literature finding a higher support towards wind turbines when the responsibility of human activities on the environment are recognized (Theron et al., 2011).

Column (3) examines the heterogeneity with respect to residential real estate prices. We use data on real estate prices in 2005 (first year available), and focus over the period 2006-2020 to avoid reverse causality issues. We label as "high real estate cost" those municipalities where the residential real estate price per square meter in 2005 was higher than the median price (i.e., roughly 575 EUR p/m$^2$). Column (3) shows that in municipalities with higher real estate costs positive reinforcement is reduced while the coefficient for center-right incumbent coalitions goes from positive to negative (although it remains non-significant). This evidence suggests that the negative impact of wind turbines on residential real estate prices (Jones and Eiser (2009), Petrova (2013)) may justify a greater sentiment of local opposition to wind projects.

Column (4) shows the results of heterogeneity analysis with respect to tourism intensity. A municipality is defined as “touristic” if it had at least one accommodation facilities in 2004 (where 50% of municipalities in southern Italy did not have an accommodation facility as of 2004). We focus on the period 2005-2020 to avoid reverse causality issues and find that being a “touristic” municipality is associated with weaker positive reinforcement sentiment for center-left incumbent coalitions and stronger electoral backlash sentiment for center-right incumbent coalitions (although the coefficient for center-right incumbent coalitions remain not statistically significant). This is in line with the qualitative results presented in Maggi et al. (2015) and based on some municipalities located in the southern region of Puglia, showing that the costs of wind turbines on tourism volumes are a concern for the local population.

**Table 5b: Incumbent coalition share on wind turbine installation: heterogeneity results - continued**

*(continue…)*

| | | (5) Plants at t-1 within municipality | | (6) Plants within 10 km | | (7) Post-2013 | | (8) 2nd half of mandate | |
|---|---|---|---|---|---|---|---|---|---|
| | | 0 | 1 | 0 | 1 | 0 | 1 | 0 | 1 |
| Coalition | C-Right | 0.49 | -0.05 | 8.58* | 7.28 | -7.93* | 0.08 | 0.19 | -9.23 |
| | | (11.46) | (13.92) | (6.86) | (7.11) | (6.54) | (4.64) | (12.60) | (15.18) |
| | C-Left | 12.56*** | 12.01 | 8.24** | 6.95*** | 2.91 | 10.92*** | 10.52*** | 17.21* |
| | | (4.78) | (11.61) | (5.05) | (3.08) | (4.22) | (3.01) | (4.59) | (9.99) |
| Observations | | 6,504 | | 6,019 | | 6,504 | | 5,334 | 5,026 |
| Years | | 2000-2020 | | 2001-2020 | | 2000-2020 | | 2000-2020 | |
| Election FE | | YES | | YES | | YES | | YES | |
| Controls | | YES | | YES | | YES | | YES | |
| 1st stage z-stat[(a)] | | 9.82 | | 10.50 | | 15.20 | | 10.08 | 8.68 |
| | | (0) | | (0) | | (0) | | (0) | (0) |

Notes: The values reported are the average treatment effects. Bootstrapped standard errors in parentheses. Significance level: *** p<0.01, ** p<0.05, * p<0.1. The estimating equation is (II) where the treatment is instrumented with the annual average wind speed 50 meters above the level of the sea in a probit-2SLS approach. Control variables include log of average income, log of population (municipality and time-varying), municipality surface and median altitude (time-invariant). The dummy “Plants at t-1 within municipality” takes value 1 if in a given municipality there were no plants at the beginning of the former mandate, and 0 otherwise. The dummy “Plants at t-1 within 10 km” takes value 1 if within 10 km from the centroid of a given municipality

there were no plants at the beginning of the former mandate, and 0 otherwise. The dummy "Post-2013" takes value 1 if t>2013 and 0 otherwise. Finally, the dummy in the last two columns "2nd half of mandate" takes value 1 if wind turbine development took place during the last two years of the previous regional government mandate, and 0 otherwise. In columns (9) the estimation of the ATE has been conducted separately using, besides the control group, treated in the first and in the second half of mandate separately. [(a)] This is the 1st stage test statistics z corresponding to the instrument's coefficient with the corresponding p-value in parentheses.

Moving to Table 5b, columns (5) and (6) provide evidence of a heterogeneous impact depending on the existence of wind turbines, either located within the municipality itself during the previous mandate (column 5), or previously located/developed within municipalities situated less than 10km from the municipality itself (column 6). In the presence of a wind turbine within the same municipality at the beginning of the former mandate positive reinforcement is slightly smaller but not in a statistically significant way. Positive reinforcement is found to be lower also in the presence of a wind turbine situated in a nearby municipality. These results seem to suggest that either the benefits from wind turbine construction are perceived less if wind projects are already present, or that there is a limit to local acceptance of wind turbines and developing further projects in areas where there are already some triggers electoral backlash.

In column (7) we test potential time heterogeneity in terms of electoral backlash and positive reinforcement. We find voters' negative reaction (i.e., electoral backlash) to be stronger in the years up to 2013 and voters' positive reaction to be stronger after 2013. This heterogeneity result may be due to the overall increase in sensitivity towards climate change over time. Moreover, as 2013 is the last year before the drop in wind turbine development observed at the national level, we may hypothesize that since 2013 municipalities where wind turbines were installed have been more carefully selected, thus reducing local opposition.

Finally, in the last column we investigate whether development during the two years prior to regional elections has a differential impact on political behavior compared to development during the first three years of the previous term.[11] We do find that the strongest reactions are observed for plants having been developed during the last two years running up to the elections, which is consistent with the hypothesis of these plants being more salient in the memory of voters.

[11] In Italy, regional elections are held every five years. In the sample comprising southern municipalities, 66% of the plants are developed during the first three years of the regional government mandate, with the remaining one third developed during the last two years.

### *4.5 Alternative treatment definitions*

Does electoral backlash or positive reinforcement depend on the size of the investment in wind turbines or public (lack of) acceptance is independent of it? To answer this question, we estimate different variants of the baseline specification. First, we substitute the treatment dummy with a dummy taking the value of 1 if at least one of the new wind turbine has a capacity greater than 1 MW and use only the municipalities where no new wind turbines were installed as a control group.[12] The results are displayed in column (1) of Table 6. The effects are similar to the baseline results. Larger plants do not seem to be associated with more intense feelings of either electoral backlash or positive reinforcement.

**Table 6: Incumbent coalition share on wind turbine installation: alternative treatment definitions**

| | (1)<br>At least one large plant | (2)<br>Increase in installed capacity |
|---|---|---|
| LATE for right incumbent | -2.68 | |
| | (7.83) | |
| LATE for left incumbent | 7.62*** | |
| | (3.94) | |
| Treat | | -3.12*** |
| | | (0.94) |
| Treat*Left | | 5.49*** |
| | | (1.02) |
| Observations | 6,250 | 6,504 |
| Election FE | YES | YES |
| Controls | YES | YES |
| Estimation Method | Probit-2SLS | IV-2SLS |
| Standard Errors | Bootstrapped | Robust |
| 1st stage Stat[(a)] | 11.74 | 35.17 |
| | (0) | |

Notes: Standard errors in parentheses. Significance level: *** p<0.01, ** p<0.05, * p<0.1. "Treat" is a dummy taking value 1 if at least a wind turbine was installed in the municipality in the previous mandate. "Left" is a dummy taking value 1 if the incumbent belong to the center-left coalition. The Probit-IV uses as instrument the annual average wind speed 50 meters above the level of the sea. Control variables include log of average income, log of population (municipality and time-varying), municipality surface and median altitude (time-invariant). The lower number of observations in the probit-IV analysis is due to the software exclusion of some observations due to failing in prediction during the first step. More in detail, all observations on year 2000 for municipalities belonging to the following regions: Abruzzo, Molise, Puglia, Basilicata, Calabria and all observations for municipalities belonging to Calabria region for year 2005. This suggests this is not a major concern. [(a)] This is the 1st stage test statistics z corresponding to the instrument's coefficient with the corresponding p-value in parentheses when the probit-2SLS is applied and the Cragg-Donald Wald F-statistics when the IV-2SLS is applied.

[12] Less than 3% of municipality-year observations feature the development of large wind turbines.

Alternatively, we report in column (2) the estimation results obtained when using total newly installed capacity in the municipality during a given mandate as treatment. When using this alternative treatment definition, a statistically significant sentiment of electoral backlash for center-right coalitions and positive reinforcement for center-left coalitions emerges. Taken together, these results suggest that the electoral backlash towards center-right coalitions is more sensitive to the size of the investment while positive reinforcement for the center-left coalitions is not.

### *4.6 External validity*

Our estimation strategy allows for the identification of a Local Average Treatment Effect (LATE), i.e., of the treatment effect in areas where wind turbines were/were not developed thanks/due to the presence/absence of high wind speed. On the other hand, it does not allow for the identification of the treatment effect where wind turbines would have been developed or would have not been developed independently from wind conditions (Angrist and Pischke, 2008).[13]

This may be an issue if local opposition, and consequently electoral backlash, in areas where wind turbines are not developed, neither in the presence nor in the absence of wind, is particularly strong, thus explaining why wind turbines in these areas are never developed, even under favorable wind conditions. If this is the case, the external validity of our findings would be impaired and the estimated effect would not describe the electoral backlash experienced in the municipalities with the highest local opposition.

In order to test for this possibility, we check whether municipalities where no wind turbine was ever developed differ from those where at least a wind turbine was developed during the period under consideration according to local characteristics that are theoretically more likely to inspire local opposition to the construction of wind plants, such as the average real estate residential price or the number of accommodation facilities. More specifically we create a dummy for absence of wind turbine installation, $\widetilde{D}_{it}$, and estimate the following probit model

$$\Phi^{-1}(\widetilde{D}_{it}) = \alpha + \lambda wind_i + \delta \boldsymbol{X}_{it} + \theta \boldsymbol{H}_{it} + \varepsilon_{it} \qquad \text{(III)}$$

where $wind_i$ is the measure of wind speed we used as an instrument in previous analysis, $\boldsymbol{X}_{it}$ is the set of controls employed in previous analysis and $\boldsymbol{H}_{it}$ includes both the residential average real estate price and the number of accommodation facilities. According to the first column in Table 7, municipalities where development never took place have a higher average real estate residential price and tourism intensity, which may be taken as suggestive evidence that the treatment effect might differ in a systematic way between municipalities where development took place and those where it did not.

[13] The possibility of results' limited external validity is a recurrent feature in frameworks were a local effect is estimated. Let us consider the case of a public subsidy to foster firms' investments in energy transition. The public subsidy is offered to the firms and some of them self-select to receive it. The researchers evaluate the impact of the subsidy on these firms' investments, estimating the average treatment effect on the treated. If the policy-makers are interested in the effect of a non-mandatory subsidy this estimation will be enough. However, if the policy-makers plan to extend the subsidy to all firms they should not base their decision on the average treatment effect on the treated alone. Indeed, the effect of the subsidy may be different on the type of firms that self-select into the treatment with respect to the type of firms that do not.

In order to investigate more accurately this possibility, we derive the set of municipalities among the “treated” ones lying on the same covariates support as the “untreated” municipalities according to a quite conservative caliper parameter (0.00001). This procedure halves the number of municipalities included in the estimation. According to a newly estimated probit model, the two groups – treated and untreated ones – are now balanced according to real estate prices and tourism intensity (column 2 of Table 7).

Based on the results from the estimation of equation (I) on the new subsample (column 3 of Table 7), there is no evidence backing up the hypothesis of a stronger electoral backlash treatment effect in municipalities where no development takes place, regardless of the presence of wind. The LATE in municipalities where the incumbent regional government was right-wing during the previous mandate is more negative compared to the baseline presented in Table 1, but not statistically significant. On the other hand, the LATE in municipalities where the incumbent regional government was left-wing is still positive, but it is no longer statistical significant. This suggests that the positive reinforcement effect of wind turbine deployment in formerly left-wing regions may be stronger in municipalities where development took place compared to those where it did not, and thus limit the external validity of our positive reinforcement results.

**Table 7: Incumbent coalition share on wind turbine installation: results external validity**

| | Absence of wind turbine installation | Incumbent coalition |
|---|---|---|

| | share | | |
|---|---|---|---|
| | (1) | (2) | (3) |
| | Balance test | Balance test after selection | Probit-IV after selection |
| Wind speed | -0.805*** | -0.493*** | |
| | (0.0698) | (0.083) | |
| Average real estate residential price | 0.000928*** | 0.000179 | |
| | (0.00021) | (0.00027) | |
| Number of accommodation facilities | 0.0115** | 0.00127 | |
| | (0.00465) | (0.00423) | |
| LATE for right incumbent | | | -1.92 |
| | | | (10.48) |
| LATE for left incumbent | | | 2.43 |
| | | | (7.64) |
| Observations | 1,681 | 876 | 876 |
| Controls | YES | YES | YES |
| Standard Errors | Robust | Robust | Bootstrapped |

Notes: Standard errors in parentheses. Significance level: *** p<0.01, ** p<0.05, * p<0.1. Control variables include log of average income, log of population (municipality and time-varying), municipality surface and median altitude (time-invariant). All variables are defined at municipality level.

## 5. Conclusions

Wind energy total installed capacity stood at 12.1 GW in 2023, corresponding to 27.5% of installed capacity in renewable energy sources (Terna, 2023). However, wind production was less than a third the one produced by the United Kingdom, and less than 20% of the German one (OurWorldInData, 2022). There is thus the impression that Italy is not making the most of its wind potential in terms of deployment of wind power generating capacity. "Not In My BackYard" (NIMBY) type of behavior related to the development of wind turbines might be hampering wind power generating capacity deployment.

In this paper, we leverage 20 years of data on wind turbine development and electoral outcomes across Italian municipalities to investigate whether a NIMBY type-of-behavior is effectively backed by the data. More specifically, we exploit variation across Italian municipalities located in the South in wind turbine development and electoral outcomes in the occasion of regional elections, provided that regions are the relevant administrative layer when it comes to legislation in the energy sector. Our approach consists of comparing the share obtained by the incumbent regional government in municipalities where development took place during the course of the previous regional government mandate to municipalities where it did not, all while instrumenting wind turbine development with wind speed.

Our main findings point in the direction of a mild and not statistically significant electoral backlash for right-wing regional administrations and of a strong and statistically significant positive reinforcement for left-wing regional administrations. The heterogeneity of the results, exclusively based on "color" rather than on the actual content of electoral programs, hints towards a high level of partisanship in the context of Italian politics. Based on these findings, we reject the hypothesis that NIMBYism is a factor hampering the deployment of wind power generating capacity in Italy.

**References**

Adams, R., H. Almeida and D. Ferreira, 2009. *Understanding the relationship between founder-CEOs and firm performance,* Journal of Empirical Finance, vol. 16, 136-150.

Angrist, J.D. and J.S. Pischke, 2008. *Mostly harmless econometrics*. Princeton University Press.

Bayulgen, O., C. Atkinson-Palombo, M. Buchanan and L. Scruggs, 2021. *Tilting at windmills? Electoral repercussions of wind turbine projects in Minnesota*, Energy Policy, vol. 159 (112636).

Bigerna, S. and P. Polinori, 2015. *Assessing the Determinants of Renewable Electricity Acceptance Integrating Meta-Analysis Regression and a Local Comprehensive Survey,* Sustainability, vol. 7, 11909-11932.

Cerulli, G., 2014. *ivtreatreg: A command for fitting binary treatment models with heterogeneous response to treatment and unobservable selection*, The Stata Journal, vol. 14 (3), 453-480.

Daniele, F., Pasquini, A., Clò, S. and Maltese, E., 2022. *Unburdening regulation: the impact of regulatory simplification on photovoltaic adoption in Italy*, Bank of Italy Working Papers n.1387.

Deiana, C. and Geraci, A., 2021. *Are wind turbines a mafia windfall? The unintended consequences of green incentives,* Regional Science and Urban Economics, vol. 89.

Devine-Wright, P., 2005. *Beyond NIMBYism: towards an Integrated Framework for Understanding Public Perceptions of Wind Energy*, Wind Energy, vol. 8, 125-139.

Ellis, G. and G. Ferraro, 2016. *The social acceptance of wind energy*, JRC Science for policy report, EUR28182.

Eurostat, 2022. Renewable energy statistics. https://ec.europa.eu/eurostat/statistics-explained/index.php?title=Renewable_energy_statistics#:~:text=In%202021%2C%20renewable%20energy%20represented,down%20from%2022.1%25%20in%202020.&text=The%20share%20of%20energy%20from,down%20from%2010.3%25%20in%202020.

Fergen, J. and J. B. Jacquet, 2016. *Beauty in motion: Expectations, attitudes, and values of wind energy development in the rural U.S.*, Energy Research & Social Science, vol. 11, 133-141.

Hoen, B., J. Firestone, J. Rand, D. Elliot, G. Hübner, J. Pohl, R. Wiser, E. Lantz, T. R. Haac, and K. Kaliski, 2019. *Attitudes of U.S. Wind Turbine Neighbors: Analysis of a Nationwide Survey*, Energy Policy, vol. 134 (110981).

Hoffmann, R., Muttarak, R., Peisker, J. and Stanig, P., 2022. *Climate change experiences raise environmental concerns and promote green voting*. Nature Climate Change.

Jones, C. R. and R. Eiser, 2009. *Identifying predictors of attitudes towards local onshore wind development with reference to an English case study*, Energy Policy, vol. 37, 4604-4614.

Jones, C. R. and R. Eiser, 2010. *Understanding 'local' opposition to wind development in the UK: How big is a backyard*, Energy Policy, vol. 38, 3106-3117.

Kontogianni, A. and Tourkolias, Ch. and Skourtos, M. and Damigos, D., 2014. *Planning globally, protesting locally: Patterns in community perceptions towards the installation of wind farms*, Renewable Energy, vol. 66, 170-177.

Maggi, M., C. Lonigro and A. Luzi, 2015. *Gli impianti eolici nella percezione di alcune comunità del Sub-Appennino Dauno*, ISPRA, Quaderni - Ambiente e Società, n° 11.

Ministero dell'Ambiente e della Sicurezza Energetica, 2023. Piano Nazionale Integrato dell'Energia e il Clima, https://www.mase.gov.it/sites/default/files/PNIEC_2023.pdf.

OurWorldinData, 2022. Renewable energy. https://ourworldindata.org/renewable-energy.

Petrova, M. A., 2013. *NIMBYism revisited: public acceptance of wind energy in the United States*, WIREs Climate Change, vol. 4, 575-601.

Stokes, L. C., 2016. *Electoral Backlash against Climate Policy: A Natural Experiment on Retrospective Voting and Local Resistance to Public Policy*, American Journal of Political Science, vol. 60 (4), 958-974.

Stratton, S., 2015. *Assessing the Accuracy of Survey Research*, Prehospital and Disaster Medicine, 30(3), 225-226.

Swofford, J. and M. Slattery, 2010. *Public attitudes of wind energy in Texas: Local communities in close proximity to wind farms and their effect on decision-making*, Energy Policy, vol. 38, 2508-2519.

Terna, 2022. Fonti rinnovabili. https://www.terna.it/it/sistema-elettrico/dispacciamento/fonti-rinnovabili.

Theron, S., J. R. Winter, D.G. Loomis and A. D. Spaulding, 2011. *Attitudes Concerning Wind Energy in Central Illinois*, Journal of the ASFMRA, vol. 2011, 1-8.

Urpelainen, J. and A. Tianbo Zhang, 2022. *Electoral Backlash or Positive Reinforcement? Wind Power and Congressional Elections in the United States*, The Journal of Politics, vol. 84 (3).

Wolsink, M., 2000. *Wind power and the NIMBY-myth: institutional capacity and the limited significance of public support*, Renewable Energy, vol. 21, 49-64.

Wooldridge, J. M., 2002. *Econometric Analysis of Cross Section and Panel Data*, The MIT Press, Cambridge, Massachusetts. London, England.

## Appendix

Robustness of the OLS results with respect to the exclusion of the following observations: all municipalities belonging to Abruzzo, Molise, Puglia, Basilicata or Calabria in 2000 and all municipalities belonging to Calabria in year 2005.

**Table A1: Incumbent coalition share on wind turbine installation: robustness check**

| | Incumbent coalition share | | | |
|---|---|---|---|---|
| | (1) | (2) | (3) | (4) |

| | OLS | OLS | OLS | OLS |
|---|---|---|---|---|
| Treat | 0.53 | -1.83 | 0.67 | -1.88 |
| | (0.56) | (1.13) | (0.56) | (1.14) |
| Treat*Left | | 3.06** | | 3.33*** |
| | | (1.30) | | (1.30) |
| Obs excluded | NO | NO | YES | YES |
| Observations | 8,152 | 8,152 | 6,505 | 6,505 |
| Macroarea | South | South | South | South |
| Election FE | YES | YES | YES | YES |
| Std errors | Robust | Robust | Robust | Bootstrapped |

Notes: Standard errors in parentheses. Significance level: *** p<0.01, ** p<0.05, * p<0.1. “Treat” is a dummy taking value 1 if at least a wind turbine was installed in the municipality in the previous mandate. “Left” is a dummy taking value 1 if the incumbent belong to the center-left coalition.